\documentclass[showpacs,amsmath,nofootinbib,amssymb, reprint,prd]{revtex4-1}
\pdfoutput=1
\usepackage{mathtools}
\mathtoolsset{showonlyrefs}
\usepackage{psfrag}
\usepackage{array}
\usepackage{amssymb}
\usepackage{amsmath}
\usepackage{amsthm}
\usepackage{graphicx}
\usepackage{epstopdf}
	
\usepackage{epsfig}
\usepackage[punctsep]{collref}
\collectsep[]{;}

\def\tN{\widetilde{N}}
\def\al{A}

\def\pb[#1,#2]{\{#1, #2\}}
\def\deb[#1,#2]{[#1,#2]_{\text{D.B.}}}

\def\tr{{\rm Tr}}

\def\ta{\widetilde{a}}

\def\Or[#1]{{\text{O}}\left({#1}\right)}
\def\dotl[#1,#2]{\left\langle #1,\, #2 \right\rangle}
\def\dotlb[#1,#2]{\left\langle #1,\, #2 \right\rangle}
\def\dotlm[#1,#2]{\left[ #1,\, #2 \right]}
\def\dotp[#1,#2]{(\vect{#1} \cdot\vect{#2})}
\def\aff[#1,#2]{\hat{#1}(#2)}

\def\nc{ S}
\def\n4sym{{\cal N}=4 SYM}
\def\>{\rangle}
\def\<{\langle}
\def\weight[#1,#2,#3]{\{(#1),#2,#3\}}
\def\ads[#1]{$\text{AdS}_{#1}$}

\def\tarelr[#1]{\widetilde{a}^{\text{rel}}_{R#1}}
\def\Oright[#1]{{\cal O}_{R#1}}
\def\Oleft[#1]{{\cal O}_{L#1}}
\def\aleft[#1]{a_{L#1}}
\def\arelr[#1]{a^{\text{rel}}_{R#1}}
\def\phirelr{{\phi}^{\text{rel,R}}}
\def\longT{T_{\text{av}}}
\def\tband{T_{\text{b}}}
\def\capt{T}
\def\cutoffT{T_{\text{cut}}}
\def\intT{T_i}
\def\EE{E'}

\def\boltzf{(1-e^{-\beta \omega_n})^{-1}}

\newcommand{\tfdT}{|\Psi_T \rangle}

\newcommand{\tfdTbra}{\langle \Psi_T |}
\newcommand{\tfdTbranol}{\langle \Psi_T}
\newcommand{\tfd}{|\Psi \rangle}

\newcommand{\be}{\begin{equation}}
\newcommand{\ee}{\end{equation}}
\newcommand{\ba}{\begin{align}}
\newcommand{\ea}{\end{align}}
\newcommand{\bs}{\begin{split}}
\def\sess\end{split}

\newcommand{\vect}[1]{{#1}}

\def\lt{t_L}
\def\rt{t_R}
\def\ang{}
\def\angp{}

\begin{document}
\title{Local Operators in the Eternal Black Hole}
\author{Kyriakos Papadodimas}
\email{kyriakos.papadodimas@cern.ch}
\affiliation{Van Swinderen Institute for Particle Physics and Gravity, University of Groningen, Nijenborgh 4, 9747 AG, The Netherlands }
\affiliation{Theory Group, Physics Department, CERN, CH-1211 Geneva 23,
Switzerland.}
\author{Suvrat Raju}
\email{suvrat@icts.res.in}
\affiliation{International Centre for Theoretical Sciences, Tata Institute of Fundamental Research, IISc Campus, Bengaluru 560012, India.}
\begin{abstract}
We show that, in the AdS/CFT correspondence, states obtained by Hamiltonian evolution of the thermofield doubled state are also dual to an eternal black hole geometry, which is glued to the boundary with a time shift generated by a large diffeomorphism. We describe gauge invariant relational observables that probe the black hole interior in these states and constrain their properties using effective field theory. By adapting recent versions of the information paradox we show that these observables are necessarily described by state-dependent bulk-boundary maps, which we construct explicitly.
\end{abstract}
\maketitle
\section{Introduction}
There has been significant recent debate on whether the AdS/CFT correspondence \cite{Maldacena:1997re,*Witten:1998qj,*Gubser:1998bc}
can meaningfully describe the interior of a black hole. The authors (AMPSS) of \cite{Almheiri:2012rt,*Almheiri:2013hfa,Marolf:2013dba} 
argued, following Mathur \cite{Mathur:2009hf}, that the CFT did not contain operators  with the right properties to play the role of local perturbative excitations in the interior of a black hole. However, in \cite{Papadodimas:2012aq, Papadodimas:2013jku,*Papadodimas:2013wnh}, we explicitly identified such operators. Our construction circumvented the AMPSS arguments 
by allowing the map between bulk and boundary operators to be state-dependent. 

In this paper, we show that versions of the AMPSS paradoxes also appear in the eternal black hole and must, once again, be resolved using state-dependent bulk-boundary maps.

It is generally accepted that the eternal black hole has a smooth interior and is 
dual to a particular entangled state of two decoupled conformal field theories \cite{Maldacena:2001kr},
\be
\label{tfddef}
\tfd = {1\over \sqrt{Z(\beta)}} \sum_E e^{-{\beta E \over 2}} | E,  E \rangle,
\ee
where the sum is over all energy eigenstates, $\beta$ is the inverse temperature of the black hole, and the partition function is  $Z(\beta) = \tr(e^{-\beta H_R})$. 

We will first argue that if one accepts this duality then it follows that a much broader class of states is dual to the same  ``geometry'', but glued differently to the boundary.  These states are obtained by evolving $\tfd$ with either of the two boundary Hamiltonians for a time $\capt$.
\be
\label{tfdmod}
\tfdT =  e^{i H_L \capt} \tfd = e^{i H_R \capt} \tfd.
\ee
Using a variant of 
the arguments of \cite{Almheiri:2013hfa,Marolf:2013dba}, we show that it is impossible to find a global linear map between bulk and boundary fields that reproduces the predictions of effective field theory behind the horizon for all states in the class \eqref{tfdmod}. 

On the other hand, if we consider a given state from \eqref{tfdmod} and small fluctuations about this state, then we can explicitly write down boundary operators that are dual to local bulk operators as we show in section \ref{secexplicit}. 

Although it is sometimes mistakenly believed \cite{Harlow:2014yoa} that state-dependent bulk-boundary maps ---  which were also explored in \cite{Verlinde:2012cy, *Verlinde:2013uja, *Verlinde:2013vja, *Verlinde:2013qya, Guica:2014dfa} --- are disallowed in quantum mechanics, we show here that they are necessary to preserve  quantum effective field theory for the infalling observer.  This suggests that state-dependence is a broader feature of local operators in quantum gravity. 
\section{Review of the Eternal Black Hole \label{secreview}}
The metric of the eternal AdS$_{d+1}$-Schwarzschild black hole is
\begin{equation}
 \label{threebrane}
 ds^2 = -f(r) dt^2 + f(r)^{-1}dr^2 + r^2 d\Omega_{d-1}^2,
\end{equation}
where $f(r) = r^2 + 1 -c_d GM r^{2-d}$, 
and $c_d = {8 (d-1)^{-1} \pi^{{2-d}\over 2} \Gamma(d/2)}$. 

We introduce the tortoise coordinate 
 ${d r_* \over d r\,\,\,} = f(r)^{-1}$ so that the right boundary is at $r_*=0$ and the 
future horizon is at $r_* \rightarrow -\infty, t \rightarrow \infty$. This metric can be smoothly extended  past the horizon by defining the Kruskal variables $U = -e^{{2 \pi\over \beta} (r_* - t)}$ and $V = e^{{2\pi  \over \beta} (r_* + t)}$. The future horizon is then at $U = 0$ and $V$ finite. The past horizon is at $V = 0$, $U$ finite. 

The quadrant connected to the right boundary is called region I. We can also introduce Schwarzschild coordinates in the other quadrants of the extended geometry. In region II, inside the black hole,
we write $U =e^{{2 \pi\over \beta} (r_* - t)}, V = e^{{2\pi \over \beta} (r_* + t)}$ and in region III, connected to the left asymptotic region,
we have $U = e^{{2\pi \over \beta } (r_* - t)}, V = -e^{{2\pi \over \beta} (r_* + t)}$. 
The two boundaries are at $UV=-1$.

It is important to note that the geometry is ``glued'' to the left CFT with a flip in the time coordinate in region III. Therefore, while the time in CFT$_R$ is identified as $\rt = t$, the time in CFT$_L$ is identified as $\lt = -t$.  Correspondingly, the isometry of the geometry,
generated by $t \rightarrow t + \capt$ in all regions of spacetime, is dual to the identity  $e^{i (H_R - H_L) \capt} \tfd = \tfd.$

\section{Time-Shifted Thermofield States \label{sectimeshifted}}
We now consider the time-shifted thermofield states defined in \eqref{tfdmod}. In any theory of quantum gravity, evolution with a boundary Hamiltonian simply corresponds to a large diffeomorphism in the bulk that does not die off at the boundary \cite{Regge:1974zd,*DeWitt:1967yk}. 

The precise action of the CFT Hamiltonian in the bulk is a gauge-dependent quantity because the Dirac brackets between the boundary Hamiltonian and bulk operators depend on the gauge-fixing conditions. However, on the boundary, its action is gauge invariant: the action of $e^{i H_L \capt}$  corresponds to a large diffeomorphism that induces a flow on the left boundary $\lt \rightarrow \lt + \capt$; correspondingly $e^{i H_R \capt}$ includes a flow on the right boundary,  $\rt \rightarrow \rt + \capt$. 

It is easy to find explicit examples of such diffeomorphisms; $e^{i H_L \capt}$ can be implemented by
\be
\label{hldiffexample}
U \rightarrow U \big[\gamma \hat{\theta}(-X) + \hat{\theta}(X) \big]; V \rightarrow {V \over \gamma} \big[\hat{\theta}(-X) + \gamma \hat{\theta}(X) \big], 
\ee
where $X=V-U$, $\gamma = e^{2 \pi \capt \over \beta}$ and $\hat{\theta}$ is a smooth version of the theta function: we set $\hat{\theta}(x) = \theta(x)$ for $|x| > \epsilon$, and the function makes a smooth transition between its values in the range $[-\epsilon, \epsilon]$ where $\epsilon \ll 1$.

In general, such a diffeomorphism changes the state. In defining the theory we do not mod out by diffeomorphisms that act non-trivially on the boundary. However, we do mod out by  trivial diffeomorphisms: diffeomorphisms with the same boundary action  --- even if they differ in the bulk ---  must be identified. The reader may find it useful to recall the Brown-Henneaux analysis \cite{Brown:1986nw} where the global AdS$_3$ vacuum was 
excited by large diffeomorphisms corresponding to the Virasoro algebra. 

Therefore the diffeomorphism in \eqref{hldiffexample} is {\em not unique}. It is a representative of an equivalence class of diffeomorphisms which all have the common property that they ``slide'' the left boundary by  $\capt$.

Since $e^{i H_L \capt}$ is just a diffeomorphism, it evidently leaves all quantities that are intrinsic to the bulk geometry invariant.
The specific statement that we will be interested in below is that an infalling observer from the {\em right} perceives a smooth horizon in all states $\tfdT$. This point was also made in the recent interesting paper \cite{Mandal:2014wfa}, and discussed in \cite{Maldacena:2013xja}.

Since we did not, anywhere, use the classical equations of motion in reaching this conclusion, and our only input was the interpretation of Hamiltonian evolution as a large diffeomorphism, we conjecture that this statement is {\em exact}. The states $\tfdT$ are smooth even for $\capt = \Or[e^{S}]$, where $S$ is the entropy of the black hole. Note that these exponentially long times are still parameterically smaller than the Poincare recurrence time \cite{Barbon:2004ce,*Barbon:2003aq}, which does not play a role in our discussion.

To support this strong claim, we provide two other perspectives. Using the isometry \eqref{tfdmod} of the thermofield state, 
the experience of the right infalling observer jumping into the state $\tfdT$ at $\rt = 0$ is the same as the experience of the right observer jumping into the state $\tfd$ at $\rt = -\capt$. From the geometry, it is clear that the right observer's experience is independent of the time at which he jumps in. If we accept this as an exact statement, then it follows that $\tfdT$  is smooth for all $\capt$.

Second, from the point of view of the CFT, we would like to treat the state $\tfdT$ on the same footing as $\tfd$ because there is no natural common origin of time in the two CFTs; the new states correspond to using a shifted time origin on the left. We explore this in more detail in \cite{longpaper}

It is also useful to recognize that the 
states $\tfdT$ are phase shifted states in the CFT.
\be
\label{tfdphase}
\tfdT = {1 \over \sqrt{Z(\beta)}} \sum_E e^{-{\beta E \over 2}} e^{i \phi[E]} | E , E \rangle,
\ee
where $\phi[E] = E \capt~\text{mod}~2 \pi$.  In fact, since the spectrum of the CFT is chaotic at high energies \cite{Festuccia:2006sa}, we can choose $\capt$ to approximate any desired phase to arbitrary accuracy for a conformal primary.
However, since the energy of conformal descendants is integrally quantized, we require that within an irreducible representation,
\be
\phi[E] - \phi[E+1] = \phi[E+1] - \phi[E+2]~\text{mod}~2 \pi.
\ee
Supersymmetric states also have integrally quantized energies but they are exponentially unimportant in \eqref{tfdphase}. 
The advantage of this perspective is that the phases corresponding to $\capt \sim \Or[1]$  are not qualitatively different from the phases generated by exponentially long $\capt$.  This makes it natural
that $\tfdT$ represents a smooth geometry even for long times.

\section{Relational Observables \label{secrelational}}
In the presence of gravity, we cannot assume that the coordinates are fixed as one changes the state. Instead we must work with gauge invariant relational observables. These are not strictly local, but behave like local operators for many purposes \cite{Giddings:2005id}. 

Intuitively, we want to start from a point on the boundary and follow a null geodesic for a given affine parameter to relationally specify a point in the bulk. To normalize the affine parameter, we must be more careful. 

First, starting  with a given boundary point $(\rt,\vect{\Omega})$, we consider a null geodesic  parameterized by ordinary asymptotic Schwarzschild time $t$, and with 
no initial velocity along $S^{d-1}: \dot{\vect{\Omega}} = 0$. We can then specify all points in front of the horizon as intersection points of this geodesic with another geodesic that hits the boundary at a later time $(\rt', \vect{\Omega'})$ with no final velocity $\dot{\vect{\Omega'}} = 0$. The value of $\vect{\Omega'}$ must be chosen to ensure these geodesics intersect. 

We can now use these points to normalize the affine parameter  \cite{longpaper} of a null geodesic, and follow it into the horizon for some affine time. 
All points in region I and region II can be reached in this manner. 
 
Now, let us consider the field $\phi$ as measured at one of these points, which we will denote by $\phirelr(\rt, \lambda)$, suppressing the dependence on the $S^{d-1}$ which is interesting \cite{longpaper} but not relevant for our discussion.  The qualifier ``rel,R'' indicates that this is a relational observable defined with respect to the right boundary.

We note that $\phirelr(\rt, \lambda)$ is invariant under any diffeomorphism that leaves the right boundary invariant {\em including} diffeomorphisms that do  not vanish at the left boundary. We can see this in several ways. By means of a trivial diffeomorphism --- one that leaves both boundaries invariant --- we can transform any diffeomorphism that induces a flow on the left boundary into one that vanishes everywhere except for a small region localized infinitesimally close to the left boundary. So, it leaves the experience of the right observer unchanged. More formally, consider a diffeomorphism taking bulk points $\vect{x} \rightarrow \vect{g}(\vect{x})$. We can equivalently represent this as an action on the  
fields $\phi(\vect{x}) \rightarrow \phi(\vect{g}^{-1}(\vect{x}))
$ and corresponding actions on the metric and higher spin fields.  On the other hand, the solution to the geodesic equation for a given affine parameter, $\lambda$, (determined as above) transforms under the new connection coefficients as $\vect{x}(\lambda) \rightarrow \vect{g}(\vect{x}(\lambda))$. So, the relational observable $\phirelr(\rt,  \lambda)$ is left invariant. 

Furthermore, diffeomorphisms that induce a flow along the right boundary just shift the value of $\rt$. Together, this implies that
\be
\label{commutH}
\begin{split}
&e^{i H_L \capt} \phirelr(\rt, \lambda) e^{-i H_L \capt} =\phirelr(\rt,\lambda), \\
&e^{i H_R \capt} \phirelr(\rt, \lambda) e^{-i H_R \capt} = \phirelr(\rt + \capt,  \lambda).
\end{split}
\ee

Now, by solving the geodesic equation in the metric given by the eternal black hole, we can trade the parameters $\rt, \lambda$ for the usual Kruskal coordinates, $U(\rt, \lambda), V(\rt, \lambda)$. Near the horizon, $U=0$, the field can be expanded in creation and annihilation operators
\begin{widetext}
 \be
\label{nearhorizonexpansionrel}
\begin{split}
&\lim_{U \rightarrow 0^-} \phirelr(U,V)  = \sum_{\omega_n}\omega_n^{-{1 \over 2}}\left[ \arelr[\omega_n \ang] \left(e^{i \delta_n} U^{i {\beta\omega_n \over 2\pi}} + e^{-i \delta_n} V^{-i {\beta\omega_n\over 2\pi}} \right) + \text{h.c} \right],\\
&\lim_{U \rightarrow 0^+} \phirelr(U,V)  = \sum_{ \omega_n}\omega_n^{-{1 \over 2}}\left[ \tarelr[\omega_n]  e^{-i \delta_n} U^{-i {\beta\omega_n\over 2\pi}} + \arelr[\omega_n \ang]  e^{-i \delta_n} V^{-i {\beta\omega_n\over 2\pi}}  + \text{h.c} \right],
\end{split}
\ee
\end{widetext}
where $\delta_n$ depends on details of scattering in the black hole geometry \cite{Papadodimas:2012aq}. The relations \eqref{commutH} now translate into
\be
\label{commutmodes}
\begin{split}
&[H_R \,,\, \arelr[\omega_n \ang]] =\, - \omega_n \,\arelr[\omega_n \ang], \quad [H_L\,, \,\arelr[\omega_n \ang]]  = 0, \\
&[H_R\,,\, \tarelr[\omega_n \ang]] = \omega_n \,\tarelr[\omega_n \ang], \quad [H_L\,,\, \tarelr[\omega_n \ang]]= 0. 
\end{split}
\ee

These surprising commutation relations show that the naive construction of local bulk operators in terms of boundary operators is incorrect. In particular, we may still identify $\arelr[\omega_n \ang]$ with the Fourier modes of an appropriate single trace operator, ${\cal O}$, in the right CFT \cite{Banks:1998dd,  *Bena:1999jv, *Hamilton:2006az, *Hamilton:2005ju, *Hamilton:2007wj, *VanRaamsdonk:2009ar, *VanRaamsdonk:2010pw,*VanRaamsdonk:2011zz}: $\arelr[\omega_n \ang] = G^{-{1 \over 2}} \Oright[\omega_n \ang]$, with
\be
\begin{split}
&\Oright[\omega_n \ang] = \tband^{-{1 \over 2}} \int_{-\tband}^{\tband} \Oright[](\rt) e^{i \omega_n \rt} d \rt, \\
&G = {1 \over Z(\beta)} \tr(e^{-\beta H_R} [\Oright[\omega_n \ang] , \Oright[\omega_n \ang]^\dagger]).
\end{split}
\ee
where $\tband$ specifies a time band that we can use to define slightly ``smeared'' Fourier modes \cite{longpaper}.

However, we cannot use  $\aleft[\omega_n \ang] = G^{-{1 \over 2}} \Oleft[\omega_n \ang]$ in place of $\tarelr[\omega_n \ang]$ in \eqref{nearhorizonexpansionrel}. This has the wrong commutator with the Hamiltonian and the wrong two-point function,
\be
\tfdTbra \aleft[\omega_n \ang] \arelr[\omega_n' \angp] \tfdT = \boltzf e^{i \omega_n \capt - { \beta \omega \over 2}} \delta_{\omega_n,\omega_n'}.
\ee
A short calculation shows that $\tarelr[\omega_n \ang]$ must have a similar two point function but without the factor of $e^{i \omega_n \capt}$; this is essential to reproduce the two point function of a field propagating about a smooth horizon.

 We now proceed to show that {\em no} state-independent operator in the CFT can play the role of $\tarelr[\omega_n \ang]$.

\section{Paradoxes in the Eternal Black Hole \label{secparadox}}
Consider the number operator as measured by the right infalling observer. With $c \equiv \arelr[{\omega_n}] - e^{-{\beta \omega_n \over 2}} (\tarelr[\omega_n])^{\dagger}$, and $d \equiv \tarelr[\omega_n] - e^{-{\beta \omega_n \over 2}} (\arelr[\omega_n])^{\dagger}$ we define 
\be
N_a = \boltzf(c^{\dagger} c + d^{\dagger} d),
\ee
At a smooth horizon the infalling observer expects to encounter no particles
\be
\tfdTbra  N_a  \tfdT = \Or[\nc^{-1}],
\ee
except for small quantum fluctuations, 
proportional to a power of the entropy $\nc$ that are also independent of time. 

If we take the long time average
\be
\begin{split}
{1 \over 2 \longT} \int_{-\longT}^{\longT} \tfdTbra N_a \tfdT dT =  \sum_E {e^{-\beta E} \over Z(\beta)} \langle E, E| N_a | E, E \rangle& \\ + 
 \sum_{E \neq \EE} {e^{-{\beta \over 2} (E + \EE) } \over Z(\beta)} {\sin{\big[(\EE - E) \longT\big]} \over (\EE - E) \longT} \langle E,E | N_a |\EE,\EE \rangle.&
\end{split}
\ee
With the averaging time, $\longT$, large enough, we see that this is only possible if 
\be
\sum_E { e^{-\beta E} \over Z(\beta)}  \langle E, E| N_a | E, E \rangle  = \Or[\nc^{-1}].
\ee
But since this is true for all $\beta$ above the Hawking-Page transition temperature, we can do a Legendre transform and conclude
\be
\langle E, E| N_a |E, E \rangle = \Or[\nc^{-1}],
\ee
for typical energy eigenstates $E$ relevant to the black hole.

Note that while we used the number operator, $N_a$, in the reasoning above,  we could have used some other operator to detect the smoothness of the horizon. The point is that if a state-independent operator predicts that the the thermofield state and its time-shifted cousins all have regular interiors, then this  operator also predicts that
eigenstate pairs of the form $|E,E\rangle$ are smooth. 

But the authors of \cite{Almheiri:2013hfa, Marolf:2013dba} argued that state-independent operators cannot describe the black hole interior in individual energy eigenstates $|E\rangle$
of a single CFT. How can the interior of energy eigenstate pairs $|E,E\rangle$
--- which do not even have any entanglement ---
be then described by such operators?

To sharpen this question, we additionally assume that the absence of entanglement implies that eigenstate pairs have no wormhole \cite{Maldacena:2013xja, *Shenker:2013pqa, *Shenker:2013yza}. Therefore no experiment on the left should affect  the right infalling observer
\be
\langle E, E | U_L^{\dagger} N_a U_L | E, E \rangle = \langle E, E | N_a | E, E \rangle, \forall U_L.
\ee
By suitably selecting $U_L$ we infer
\be
\label{eigenstatesmooth}
\langle \EE, E | N_a  | \EE, E \rangle = \Or[\nc^{-1}],
\ee
where $\EE$ is an {\em independent} fixed typical energy eigenstate in the left CFT.

We now immediately run into the AMPSS paradoxes. For example, if we denote the orthonormal eigenstates of the number operator $a_{\omega_n}^{\dagger} a_{\omega_n}$ by $|N_j \rangle$, then we expect 
\be
\label{schfirewalls}
\langle \EE, N_j | N_a  | \EE, N_j \rangle = \Or[1].
\ee
This is inconsistent with \eqref{eigenstatesmooth} by a change of basis \cite{Marolf:2013dba}.

To see another paradox, consider a band of energies $(E-\Delta, E+ \Delta)$ that contains ${\cal D}$ eigenstates; let $P_{E,\Delta,R}$ and $P_{E,\Delta, L}$ be the projector onto this subspace in the right and left CFTs respective. With $\tN_n \equiv \ta_{\omega_n \ang}^{\dagger}  \ta_{\omega_n \ang}$,
\be
\tr(P_{E,\Delta,L} P_{E,\Delta,R} \tN_n) =  \tr( P_{E,\Delta,L} P_{E+\omega_n,\Delta,R} ( \tN_n + 1)),
\ee
where we have used the cyclicity of the trace, and the expected commutator of the mirror operator with its adjoint and with the two Hamiltonians.
The number of states in the shifted band 
$E + \omega_n \pm \Delta$ is ${\cal D} e^{\beta \omega_n}$. A little algebra leads to
\be
\langle \EE, E | \tN_n  | \EE , E \rangle = - \boltzf. \ee
This negative expectation for a manifestly positive operator is absurd
and suggests that there is no state-independent operator $\ta_{\omega_n}$ with the expected commutation relations.

\section{State Dependent Construction of the Interior \label{secexplicit}}
To correctly construct the interior of the eternal black hole, we must use state-dependent operators, and drop the requirement that the  {\em same} operators describe the right relational observables in all time-shifted states.  These operators can be obtained as a solution to the linear
equations presented in \cite{Papadodimas:2013jku,*Papadodimas:2013wnh}. We present this solution below.

Consider the space of  states formed by exciting a given time-shifted state by operators $A_{\alpha}$ that can be written as polynomials  comprising $\Or[1]$ products of single trace operators. This space is defined more precisely in  \cite{Papadodimas:2013jku,*Papadodimas:2013wnh}. We write the projector onto this space as $\hat{P}_{\capt}$. \be
\hat{P}_{\capt} \al_{\alpha}\tfdT = \al_{\alpha} \tfdT;~ \langle v | \al_{\alpha} \tfdT = 0, \forall \alpha \Rightarrow \hat{P}_{\capt} | v \rangle = 0.
\ee

The time-shifted states are mutually almost orthogonal
\be
\label{innerproduct}
|\tfdTbranol \tfd |^2 = \left(1 + \Or[S^{-1}] \right) e^{-{\capt^2 C \over \beta^2}},
\ee
where $C \propto \nc$ is the specific heat of the CFT,  and the expression is valid for $\capt \ll 1$.

So, by means of an  $\Or[1]$ cutoff, $\cutoffT$,  we construct
\be
\label{statedependental}
\tarelr[\omega_n \ang] = {\sqrt{ C \over \pi \beta^2}} \int_{\capt-\cutoffT}^{\capt+\cutoffT } a^{\intT}_{L \omega_n \ang} \hat{P}_{\intT} d \intT,
\ee
where 
\be
a^{\intT}_{L \omega_n \ang} = \tband^{-{1 \over 2}} 
\int_{-\tband}^{\tband} \Oleft[](\intT + \lt) e^{i \omega_n \lt} d\lt
\ee
 is a smeared Fourier mode of of the left field evaluated about $\intT$. When inserted in correlators about the state $\tfdT$, and in neighbouring states obtained by acting with ``reasonable'' excitations (including $e^{i H \intT}$ for $\intT < \cutoffT$)  it is easy to check that $\tarelr[\omega_n \ang]$ satisfies the properties that we need up to $\Or[S^{-1}]$ corrections \cite{longpaper}.
 
On the other hand, for large $\capt$ the inner-product \eqref{innerproduct} does not decrease indefinitely but saturates at $\Or[e^{-S/2}]$.
If we attempt to take the cutoff $\cutoffT$ to be exponentially large, then the ``fat tail'' from the inner product above implies that interference from distant microstates spoils the good properties of the operator. 
In particular, therefore, we cannot use the same operator to describe the interior of the black hole in the entire range of states given by \eqref{tfdmod}.

\section{Conclusions}
From the point of view of the right observer, the different time-shifted states  are like ``microstates'' of the eternal black hole. In this paper, we have shown that it is impossible to construct state-independent operators that describe a smooth interior in all these microstates.
As in our previous paper, given a particular time-shifted state, we can find 
state-dependent operators with the right properties in reasonable excitations of this state. 

There are two alternatives. First, one may declare that the thermofield state itself lacks a smooth horizon \cite{Mathur:2014dia,*Avery:2013bea}. However, this contradicts AdS/CFT calculations that appear to explicitly probe the interior \cite{Kraus:2002iv,*Fidkowski:2003nf,*Hartman:2013qma}. Or one may posit that the state develops a firewall after some long time. We presented several arguments to the contrary, but the strongest is that we can explicitly construct the interior using \eqref{statedependental} even for late times. 

It is an unusual, and very interesting aspect of local operators in quantum gravity that one can prove that the ``same observable'', like a field at a ``given'' point in space, is described by different operators in different patches of the Hilbert space. This striking feature
deserves further attention.

\section*{Acknowledgments}
We have discussed these ideas with a large number of people over the past year. We are particularly grateful to all members of the string theory groups at ICTS-TIFR, TIFR (Mumbai),  IISc (Bangalore), CERN, and the University of Groningen.
We are  also grateful to the participants in the summer workshop on ``Emergent Spacetime in String Theory'' (2014) at the Aspen Center for Physics, which was supported by NSF Grant No. PHYS-1066293 and the Simons Foundation, and the Santa Barbara Gravity Workshop II (2014).
S.R. is partially supported by a Ramanujan fellowship of the Department of Science and Technology (India). K.P. would like to thank
the Royal Netherlands Academy of Sciences (KNAW).
\bibliography{references}

\begin{thebibliography}{42}%
\makeatletter
\providecommand \@ifxundefined [1]{%
 \@ifx{#1\undefined}
}%
\providecommand \@ifnum [1]{%
 \ifnum #1\expandafter \@firstoftwo
 \else \expandafter \@secondoftwo
 \fi
}%
\providecommand \@ifx [1]{%
 \ifx #1\expandafter \@firstoftwo
 \else \expandafter \@secondoftwo
 \fi
}%
\providecommand \natexlab [1]{#1}%
\providecommand \enquote  [1]{``#1''}%
\providecommand \bibnamefont  [1]{#1}%
\providecommand \bibfnamefont [1]{#1}%
\providecommand \citenamefont [1]{#1}%
\providecommand \href@noop [0]{\@secondoftwo}%
\providecommand \href [0]{\begingroup \@sanitize@url \@href}%
\providecommand \@href[1]{\@@startlink{#1}\@@href}%
\providecommand \@@href[1]{\endgroup#1\@@endlink}%
\providecommand \@sanitize@url [0]{\catcode `\\12\catcode `\$12\catcode
  `\&12\catcode `\#12\catcode `\^12\catcode `\_12\catcode `\%12\relax}%
\providecommand \@@startlink[1]{}%
\providecommand \@@endlink[0]{}%
\providecommand \url  [0]{\begingroup\@sanitize@url \@url }%
\providecommand \@url [1]{\endgroup\@href {#1}{\urlprefix }}%
\providecommand \urlprefix  [0]{URL }%
\providecommand \Eprint [0]{\href }%
\providecommand \doibase [0]{http://dx.doi.org/}%
\providecommand \selectlanguage [0]{\@gobble}%
\providecommand \bibinfo  [0]{\@secondoftwo}%
\providecommand \bibfield  [0]{\@secondoftwo}%
\providecommand \translation [1]{[#1]}%
\providecommand \BibitemOpen [0]{}%
\providecommand \bibitemStop [0]{}%
\providecommand \bibitemNoStop [0]{.\EOS\space}%
\providecommand \EOS [0]{\spacefactor3000\relax}%
\providecommand \BibitemShut  [1]{\csname bibitem#1\endcsname}%
\let\auto@bib@innerbib\@empty
%</preamble>
\bibitem [{\citenamefont {Maldacena}(1998)}]{Maldacena:1997re}%
  \BibitemOpen
  \bibfield  {author} {\bibinfo {author} {\bibfnamefont {J.~M.}\ \bibnamefont
  {Maldacena}},\ }\href {\doibase 10.1023/A:1026654312961} {\bibfield
  {journal} {\bibinfo  {journal} {Adv.Theor.Math.Phys.}\ }\textbf {\bibinfo
  {volume} {2}},\ \bibinfo {pages} {231} (\bibinfo {year} {1998})},\ \Eprint
  {http://arxiv.org/abs/hep-th/9711200} {arXiv:hep-th/9711200} \BibitemShut
  {NoStop}%
\bibitem [{\citenamefont {Witten}(1998)}]{Witten:1998qj}%
  \BibitemOpen
  \bibfield  {author} {\bibinfo {author} {\bibfnamefont {E.}~\bibnamefont
  {Witten}},\ }\href@noop {} {\bibfield  {journal} {\bibinfo  {journal} {Adv.
  Theor. Math. Phys.}\ }\textbf {\bibinfo {volume} {2}},\ \bibinfo {pages}
  {253} (\bibinfo {year} {1998})},\ \Eprint
  {http://arxiv.org/abs/hep-th/9802150} {arXiv:hep-th/9802150} \BibitemShut
  {NoStop}%
%%CITATION = HEP-TH/9802150;%%
\bibitem [{\citenamefont {Gubser}\ \emph {et~al.}(1998)\citenamefont {Gubser},
  \citenamefont {Klebanov},\ and\ \citenamefont {Polyakov}}]{Gubser:1998bc}%
  \BibitemOpen
  \bibfield  {author} {\bibinfo {author} {\bibfnamefont {S.}~\bibnamefont
  {Gubser}}, \bibinfo {author} {\bibfnamefont {I.~R.}\ \bibnamefont
  {Klebanov}}, \ and\ \bibinfo {author} {\bibfnamefont {A.~M.}\ \bibnamefont
  {Polyakov}},\ }\href {\doibase 10.1016/S0370-2693(98)00377-3} {\bibfield
  {journal} {\bibinfo  {journal} {Phys.Lett.}\ }\textbf {\bibinfo {volume}
  {B428}},\ \bibinfo {pages} {105} (\bibinfo {year} {1998})},\ \Eprint
  {http://arxiv.org/abs/hep-th/9802109} {arXiv:hep-th/9802109} \BibitemShut
  {NoStop}%
%%CITATION = HEP-TH/9802109;%%
\bibitem [{\citenamefont {Almheiri}\ \emph
  {et~al.}(2013{\natexlab{a}})\citenamefont {Almheiri}, \citenamefont {Marolf},
  \citenamefont {Polchinski},\ and\ \citenamefont {Sully}}]{Almheiri:2012rt}%
  \BibitemOpen
  \bibfield  {author} {\bibinfo {author} {\bibfnamefont {A.}~\bibnamefont
  {Almheiri}}, \bibinfo {author} {\bibfnamefont {D.}~\bibnamefont {Marolf}},
  \bibinfo {author} {\bibfnamefont {J.}~\bibnamefont {Polchinski}}, \ and\
  \bibinfo {author} {\bibfnamefont {J.}~\bibnamefont {Sully}},\ }\href
  {\doibase 10.1007/JHEP02(2013)062} {\bibfield  {journal} {\bibinfo  {journal}
  {JHEP}\ }\textbf {\bibinfo {volume} {1302}},\ \bibinfo {pages} {062}
  (\bibinfo {year} {2013}{\natexlab{a}})},\ \Eprint
  {http://arxiv.org/abs/1207.3123} {arXiv:1207.3123} \BibitemShut {NoStop}%
%%CITATION = ARXIV:1207.3123;%%
\bibitem [{\citenamefont {Almheiri}\ \emph
  {et~al.}(2013{\natexlab{b}})\citenamefont {Almheiri}, \citenamefont {Marolf},
  \citenamefont {Polchinski}, \citenamefont {Stanford},\ and\ \citenamefont
  {Sully}}]{Almheiri:2013hfa}%
  \BibitemOpen
  \bibfield  {author} {\bibinfo {author} {\bibfnamefont {A.}~\bibnamefont
  {Almheiri}}, \bibinfo {author} {\bibfnamefont {D.}~\bibnamefont {Marolf}},
  \bibinfo {author} {\bibfnamefont {J.}~\bibnamefont {Polchinski}}, \bibinfo
  {author} {\bibfnamefont {D.}~\bibnamefont {Stanford}}, \ and\ \bibinfo
  {author} {\bibfnamefont {J.}~\bibnamefont {Sully}},\ }\href {\doibase
  10.1007/JHEP09(2013)018} {\bibfield  {journal} {\bibinfo  {journal} {JHEP}\
  }\textbf {\bibinfo {volume} {1309}},\ \bibinfo {pages} {018} (\bibinfo {year}
  {2013}{\natexlab{b}})},\ \Eprint {http://arxiv.org/abs/1304.6483}
  {arXiv:1304.6483} \BibitemShut {NoStop}%
%%CITATION = ARXIV:1304.6483;%%
\bibitem [{\citenamefont {Marolf}\ and\ \citenamefont
  {Polchinski}(2013)}]{Marolf:2013dba}%
  \BibitemOpen
  \bibfield  {author} {\bibinfo {author} {\bibfnamefont {D.}~\bibnamefont
  {Marolf}}\ and\ \bibinfo {author} {\bibfnamefont {J.}~\bibnamefont
  {Polchinski}},\ }\href {\doibase 10.1103/PhysRevLett.111.171301} {\bibfield
  {journal} {\bibinfo  {journal} {Phys.Rev.Lett.}\ }\textbf {\bibinfo {volume}
  {111}},\ \bibinfo {pages} {171301} (\bibinfo {year} {2013})},\ \Eprint
  {http://arxiv.org/abs/1307.4706} {arXiv:1307.4706} \BibitemShut {NoStop}%
%%CITATION = ARXIV:1307.4706;%%
\bibitem [{\citenamefont {Mathur}(2009)}]{Mathur:2009hf}%
  \BibitemOpen
  \bibfield  {author} {\bibinfo {author} {\bibfnamefont {S.~D.}\ \bibnamefont
  {Mathur}},\ }\href {\doibase 10.1088/0264-9381/26/22/224001} {\bibfield
  {journal} {\bibinfo  {journal} {Class.Quant.Grav.}\ }\textbf {\bibinfo
  {volume} {26}},\ \bibinfo {pages} {224001} (\bibinfo {year} {2009})},\
  \Eprint {http://arxiv.org/abs/0909.1038} {arXiv:0909.1038} \BibitemShut
  {NoStop}%
%%CITATION = ARXIV:0909.1038;%%
\bibitem [{\citenamefont {Papadodimas}\ and\ \citenamefont
  {Raju}(2013)}]{Papadodimas:2012aq}%
  \BibitemOpen
  \bibfield  {author} {\bibinfo {author} {\bibfnamefont {K.}~\bibnamefont
  {Papadodimas}}\ and\ \bibinfo {author} {\bibfnamefont {S.}~\bibnamefont
  {Raju}},\ }\href {\doibase 10.1007/JHEP10(2013)212} {\bibfield  {journal}
  {\bibinfo  {journal} {JHEP}\ }\textbf {\bibinfo {volume} {1310}},\ \bibinfo
  {pages} {212} (\bibinfo {year} {2013})},\ \Eprint
  {http://arxiv.org/abs/1211.6767} {arXiv:1211.6767} \BibitemShut {NoStop}%
%%CITATION = ARXIV:1211.6767;%%
\bibitem [{\citenamefont {Papadodimas}\ and\ \citenamefont
  {Raju}(2014{\natexlab{a}})}]{Papadodimas:2013jku}%
  \BibitemOpen
  \bibfield  {author} {\bibinfo {author} {\bibfnamefont {K.}~\bibnamefont
  {Papadodimas}}\ and\ \bibinfo {author} {\bibfnamefont {S.}~\bibnamefont
  {Raju}},\ }\href {\doibase 10.1103/PhysRevD.89.086010} {\bibfield  {journal}
  {\bibinfo  {journal} {Phys.Rev.}\ }\textbf {\bibinfo {volume} {D89}},\
  \bibinfo {pages} {086010} (\bibinfo {year} {2014}{\natexlab{a}})},\ \Eprint
  {http://arxiv.org/abs/1310.6335} {arXiv:1310.6335} \BibitemShut {NoStop}%
%%CITATION = ARXIV:1310.6335;%%
\bibitem [{\citenamefont {Papadodimas}\ and\ \citenamefont
  {Raju}(2014{\natexlab{b}})}]{Papadodimas:2013wnh}%
  \BibitemOpen
  \bibfield  {author} {\bibinfo {author} {\bibfnamefont {K.}~\bibnamefont
  {Papadodimas}}\ and\ \bibinfo {author} {\bibfnamefont {S.}~\bibnamefont
  {Raju}},\ }\href {\doibase 10.1103/PhysRevLett.112.051301} {\bibfield
  {journal} {\bibinfo  {journal} {Phys.Rev.Lett.}\ }\textbf {\bibinfo {volume}
  {112}},\ \bibinfo {pages} {051301} (\bibinfo {year} {2014}{\natexlab{b}})},\
  \Eprint {http://arxiv.org/abs/1310.6334} {arXiv:1310.6334} \BibitemShut
  {NoStop}%
%%CITATION = ARXIV:1310.6334;%%
\bibitem [{\citenamefont {Maldacena}(2003)}]{Maldacena:2001kr}%
  \BibitemOpen
  \bibfield  {author} {\bibinfo {author} {\bibfnamefont {J.~M.}\ \bibnamefont
  {Maldacena}},\ }\href@noop {} {\bibfield  {journal} {\bibinfo  {journal}
  {JHEP}\ }\textbf {\bibinfo {volume} {0304}},\ \bibinfo {pages} {021}
  (\bibinfo {year} {2003})},\ \Eprint {http://arxiv.org/abs/hep-th/0106112}
  {arXiv:hep-th/0106112} \BibitemShut {NoStop}%
%%CITATION = HEP-TH/0106112;%%
\bibitem [{\citenamefont {Harlow}(2014)}]{Harlow:2014yoa}%
  \BibitemOpen
  \bibfield  {author} {\bibinfo {author} {\bibfnamefont {D.}~\bibnamefont
  {Harlow}},\ }\href {\doibase 10.1007/JHEP11(2014)055} {\bibfield  {journal}
  {\bibinfo  {journal} {JHEP}\ }\textbf {\bibinfo {volume} {1411}},\ \bibinfo
  {pages} {055} (\bibinfo {year} {2014})},\ \Eprint
  {http://arxiv.org/abs/1405.1995} {arXiv:1405.1995} \BibitemShut {NoStop}%
%%CITATION = ARXIV:1405.1995;%%
\bibitem [{\citenamefont {Verlinde}\ and\ \citenamefont
  {Verlinde}(2013{\natexlab{a}})}]{Verlinde:2012cy}%
  \BibitemOpen
  \bibfield  {author} {\bibinfo {author} {\bibfnamefont {E.}~\bibnamefont
  {Verlinde}}\ and\ \bibinfo {author} {\bibfnamefont {H.}~\bibnamefont
  {Verlinde}},\ }\href {\doibase 10.1007/JHEP10(2013)107} {\bibfield  {journal}
  {\bibinfo  {journal} {JHEP}\ }\textbf {\bibinfo {volume} {1310}},\ \bibinfo
  {pages} {107} (\bibinfo {year} {2013}{\natexlab{a}})},\ \Eprint
  {http://arxiv.org/abs/1211.6913} {arXiv:1211.6913} \BibitemShut {NoStop}%
%%CITATION = ARXIV:1211.6913;%%
\bibitem [{\citenamefont {Verlinde}\ and\ \citenamefont
  {Verlinde}(2013{\natexlab{b}})}]{Verlinde:2013uja}%
  \BibitemOpen
  \bibfield  {author} {\bibinfo {author} {\bibfnamefont {E.}~\bibnamefont
  {Verlinde}}\ and\ \bibinfo {author} {\bibfnamefont {H.}~\bibnamefont
  {Verlinde}},\ }\href@noop {} {\  (\bibinfo {year} {2013}{\natexlab{b}})},\
  \Eprint {http://arxiv.org/abs/1306.0515} {arXiv:1306.0515} \BibitemShut
  {NoStop}%
%%CITATION = ARXIV:1306.0515;%%
\bibitem [{\citenamefont {Verlinde}\ and\ \citenamefont
  {Verlinde}(2013{\natexlab{c}})}]{Verlinde:2013vja}%
  \BibitemOpen
  \bibfield  {author} {\bibinfo {author} {\bibfnamefont {E.}~\bibnamefont
  {Verlinde}}\ and\ \bibinfo {author} {\bibfnamefont {H.}~\bibnamefont
  {Verlinde}},\ }\href@noop {} {\  (\bibinfo {year} {2013}{\natexlab{c}})},\
  \Eprint {http://arxiv.org/abs/1306.0516} {arXiv:1306.0516} \BibitemShut
  {NoStop}%
%%CITATION = ARXIV:1306.0516;%%
\bibitem [{\citenamefont {Verlinde}\ and\ \citenamefont
  {Verlinde}(2013{\natexlab{d}})}]{Verlinde:2013qya}%
  \BibitemOpen
  \bibfield  {author} {\bibinfo {author} {\bibfnamefont {E.}~\bibnamefont
  {Verlinde}}\ and\ \bibinfo {author} {\bibfnamefont {H.}~\bibnamefont
  {Verlinde}},\ }\href@noop {} {\  (\bibinfo {year} {2013}{\natexlab{d}})},\
  \Eprint {http://arxiv.org/abs/1311.1137} {arXiv:1311.1137} \BibitemShut
  {NoStop}%
%%CITATION = ARXIV:1311.1137;%%
\bibitem [{\citenamefont {Guica}\ and\ \citenamefont
  {Ross}(2014)}]{Guica:2014dfa}%
  \BibitemOpen
  \bibfield  {author} {\bibinfo {author} {\bibfnamefont {M.}~\bibnamefont
  {Guica}}\ and\ \bibinfo {author} {\bibfnamefont {S.~F.}\ \bibnamefont
  {Ross}},\ }\href {\doibase 10.1088/0264-9381/32/5/055014} {\  (\bibinfo
  {year} {2014}),\ 10.1088/0264-9381/32/5/055014},\ \Eprint
  {http://arxiv.org/abs/1412.1084} {arXiv:1412.1084} \BibitemShut {NoStop}%
%%CITATION = ARXIV:1412.1084;%%
\bibitem [{\citenamefont {Regge}\ and\ \citenamefont
  {Teitelboim}(1974)}]{Regge:1974zd}%
  \BibitemOpen
  \bibfield  {author} {\bibinfo {author} {\bibfnamefont {T.}~\bibnamefont
  {Regge}}\ and\ \bibinfo {author} {\bibfnamefont {C.}~\bibnamefont
  {Teitelboim}},\ }\href {\doibase 10.1016/0003-4916(74)90404-7} {\bibfield
  {journal} {\bibinfo  {journal} {Annals Phys.}\ }\textbf {\bibinfo {volume}
  {88}},\ \bibinfo {pages} {286} (\bibinfo {year} {1974})}\BibitemShut
  {NoStop}%
%%CITATION = APNYA,88,286;%%
\bibitem [{\citenamefont {DeWitt}(1967)}]{DeWitt:1967yk}%
  \BibitemOpen
  \bibfield  {author} {\bibinfo {author} {\bibfnamefont {B.~S.}\ \bibnamefont
  {DeWitt}},\ }\href {\doibase 10.1103/PhysRev.160.1113} {\bibfield  {journal}
  {\bibinfo  {journal} {Phys.Rev.}\ }\textbf {\bibinfo {volume} {160}},\
  \bibinfo {pages} {1113} (\bibinfo {year} {1967})}\BibitemShut {NoStop}%
%%CITATION = PHRVA,160,1113;%%
\bibitem [{\citenamefont {Brown}\ and\ \citenamefont
  {Henneaux}(1986)}]{Brown:1986nw}%
  \BibitemOpen
  \bibfield  {author} {\bibinfo {author} {\bibfnamefont {J.~D.}\ \bibnamefont
  {Brown}}\ and\ \bibinfo {author} {\bibfnamefont {M.}~\bibnamefont
  {Henneaux}},\ }\href {\doibase 10.1007/BF01211590} {\bibfield  {journal}
  {\bibinfo  {journal} {Commun.Math.Phys.}\ }\textbf {\bibinfo {volume}
  {104}},\ \bibinfo {pages} {207} (\bibinfo {year} {1986})}\BibitemShut
  {NoStop}%
%%CITATION = CMPHA,104,207;%%
\bibitem [{\citenamefont {Mandal}\ \emph {et~al.}(2015)\citenamefont {Mandal},
  \citenamefont {Sinha},\ and\ \citenamefont {Sorokhaibam}}]{Mandal:2014wfa}%
  \BibitemOpen
  \bibfield  {author} {\bibinfo {author} {\bibfnamefont {G.}~\bibnamefont
  {Mandal}}, \bibinfo {author} {\bibfnamefont {R.}~\bibnamefont {Sinha}}, \
  and\ \bibinfo {author} {\bibfnamefont {N.}~\bibnamefont {Sorokhaibam}},\
  }\href {\doibase 10.1007/JHEP01(2015)036} {\bibfield  {journal} {\bibinfo
  {journal} {JHEP}\ }\textbf {\bibinfo {volume} {1501}},\ \bibinfo {pages}
  {036} (\bibinfo {year} {2015})},\ \Eprint {http://arxiv.org/abs/1405.6695}
  {arXiv:1405.6695} \BibitemShut {NoStop}%
%%CITATION = ARXIV:1405.6695;%%
\bibitem [{\citenamefont {Maldacena}\ and\ \citenamefont
  {Susskind}(2013)}]{Maldacena:2013xja}%
  \BibitemOpen
  \bibfield  {author} {\bibinfo {author} {\bibfnamefont {J.}~\bibnamefont
  {Maldacena}}\ and\ \bibinfo {author} {\bibfnamefont {L.}~\bibnamefont
  {Susskind}},\ }\href {\doibase 10.1002/prop.201300020} {\bibfield  {journal}
  {\bibinfo  {journal} {Fortsch.Phys.}\ }\textbf {\bibinfo {volume} {61}},\
  \bibinfo {pages} {781} (\bibinfo {year} {2013})},\ \Eprint
  {http://arxiv.org/abs/1306.0533} {arXiv:1306.0533} \BibitemShut {NoStop}%
%%CITATION = ARXIV:1306.0533;%%
\bibitem [{\citenamefont {Barbon}\ and\ \citenamefont
  {Rabinovici}(2004)}]{Barbon:2004ce}%
  \BibitemOpen
  \bibfield  {author} {\bibinfo {author} {\bibfnamefont {J.}~\bibnamefont
  {Barbon}}\ and\ \bibinfo {author} {\bibfnamefont {E.}~\bibnamefont
  {Rabinovici}},\ }\href {\doibase 10.1002/prop.200410157} {\bibfield
  {journal} {\bibinfo  {journal} {Fortsch.Phys.}\ }\textbf {\bibinfo {volume}
  {52}},\ \bibinfo {pages} {642} (\bibinfo {year} {2004})},\ \Eprint
  {http://arxiv.org/abs/hep-th/0403268} {arXiv:hep-th/0403268} \BibitemShut
  {NoStop}%
%%CITATION = HEP-TH/0403268;%%
\bibitem [{\citenamefont {Barbon}\ and\ \citenamefont
  {Rabinovici}(2003)}]{Barbon:2003aq}%
  \BibitemOpen
  \bibfield  {author} {\bibinfo {author} {\bibfnamefont {J.}~\bibnamefont
  {Barbon}}\ and\ \bibinfo {author} {\bibfnamefont {E.}~\bibnamefont
  {Rabinovici}},\ }\href {\doibase 10.1088/1126-6708/2003/11/047} {\bibfield
  {journal} {\bibinfo  {journal} {JHEP}\ }\textbf {\bibinfo {volume} {0311}},\
  \bibinfo {pages} {047} (\bibinfo {year} {2003})},\ \Eprint
  {http://arxiv.org/abs/hep-th/0308063} {arXiv:hep-th/0308063} \BibitemShut
  {NoStop}%
%%CITATION = HEP-TH/0308063;%%
\bibitem [{\citenamefont {Papadodimas}\ and\ \citenamefont
  {Raju}(2015)}]{longpaper}%
  \BibitemOpen
  \bibfield  {author} {\bibinfo {author} {\bibfnamefont {K.}~\bibnamefont
  {Papadodimas}}\ and\ \bibinfo {author} {\bibfnamefont {S.}~\bibnamefont
  {Raju}},\ }\href@noop {} {\bibfield  {journal} {\bibinfo  {journal} {to
  appear}\ } (\bibinfo {year} {2015})}\BibitemShut {NoStop}%
%%CITATION = ARXIV:1310.6334;%%
\bibitem [{\citenamefont {Festuccia}\ and\ \citenamefont
  {Liu}(2007)}]{Festuccia:2006sa}%
  \BibitemOpen
  \bibfield  {author} {\bibinfo {author} {\bibfnamefont {G.}~\bibnamefont
  {Festuccia}}\ and\ \bibinfo {author} {\bibfnamefont {H.}~\bibnamefont
  {Liu}},\ }\href {\doibase 10.1088/1126-6708/2007/12/027} {\bibfield
  {journal} {\bibinfo  {journal} {JHEP}\ }\textbf {\bibinfo {volume} {0712}},\
  \bibinfo {pages} {027} (\bibinfo {year} {2007})},\ \Eprint
  {http://arxiv.org/abs/hep-th/0611098} {arXiv:hep-th/0611098} \BibitemShut
  {NoStop}%
%%CITATION = HEP-TH/0611098;%%
\bibitem [{\citenamefont {Giddings}\ \emph {et~al.}(2006)\citenamefont
  {Giddings}, \citenamefont {Marolf},\ and\ \citenamefont
  {Hartle}}]{Giddings:2005id}%
  \BibitemOpen
  \bibfield  {author} {\bibinfo {author} {\bibfnamefont {S.~B.}\ \bibnamefont
  {Giddings}}, \bibinfo {author} {\bibfnamefont {D.}~\bibnamefont {Marolf}}, \
  and\ \bibinfo {author} {\bibfnamefont {J.~B.}\ \bibnamefont {Hartle}},\
  }\href {\doibase 10.1103/PhysRevD.74.064018} {\bibfield  {journal} {\bibinfo
  {journal} {Phys.Rev.}\ }\textbf {\bibinfo {volume} {D74}},\ \bibinfo {pages}
  {064018} (\bibinfo {year} {2006})},\ \Eprint
  {http://arxiv.org/abs/hep-th/0512200} {arXiv:hep-th/0512200} \BibitemShut
  {NoStop}%
%%CITATION = HEP-TH/0512200;%%
\bibitem [{\citenamefont {Banks}\ \emph {et~al.}(1998)\citenamefont {Banks},
  \citenamefont {Douglas}, \citenamefont {Horowitz},\ and\ \citenamefont
  {Martinec}}]{Banks:1998dd}%
  \BibitemOpen
  \bibfield  {author} {\bibinfo {author} {\bibfnamefont {T.}~\bibnamefont
  {Banks}}, \bibinfo {author} {\bibfnamefont {M.~R.}\ \bibnamefont {Douglas}},
  \bibinfo {author} {\bibfnamefont {G.~T.}\ \bibnamefont {Horowitz}}, \ and\
  \bibinfo {author} {\bibfnamefont {E.~J.}\ \bibnamefont {Martinec}},\
  }\href@noop {} {\  (\bibinfo {year} {1998})},\ \Eprint
  {http://arxiv.org/abs/hep-th/9808016} {arXiv:hep-th/9808016} \BibitemShut
  {NoStop}%
%%CITATION = HEP-TH/9808016;%%
\bibitem [{\citenamefont {Bena}(2000)}]{Bena:1999jv}%
  \BibitemOpen
  \bibfield  {author} {\bibinfo {author} {\bibfnamefont {I.}~\bibnamefont
  {Bena}},\ }\href {\doibase 10.1103/PhysRevD.62.066007} {\bibfield  {journal}
  {\bibinfo  {journal} {Phys.Rev.}\ }\textbf {\bibinfo {volume} {D62}},\
  \bibinfo {pages} {066007} (\bibinfo {year} {2000})},\ \Eprint
  {http://arxiv.org/abs/hep-th/9905186} {arXiv:hep-th/9905186} \BibitemShut
  {NoStop}%
%%CITATION = HEP-TH/9905186;%%
\bibitem [{\citenamefont {Hamilton}\ \emph
  {et~al.}(2006{\natexlab{a}})\citenamefont {Hamilton}, \citenamefont {Kabat},
  \citenamefont {Lifschytz},\ and\ \citenamefont {Lowe}}]{Hamilton:2006az}%
  \BibitemOpen
  \bibfield  {author} {\bibinfo {author} {\bibfnamefont {A.}~\bibnamefont
  {Hamilton}}, \bibinfo {author} {\bibfnamefont {D.~N.}\ \bibnamefont {Kabat}},
  \bibinfo {author} {\bibfnamefont {G.}~\bibnamefont {Lifschytz}}, \ and\
  \bibinfo {author} {\bibfnamefont {D.~A.}\ \bibnamefont {Lowe}},\ }\href
  {\doibase 10.1103/PhysRevD.74.066009} {\bibfield  {journal} {\bibinfo
  {journal} {Phys.Rev.}\ }\textbf {\bibinfo {volume} {D74}},\ \bibinfo {pages}
  {066009} (\bibinfo {year} {2006}{\natexlab{a}})},\ \Eprint
  {http://arxiv.org/abs/hep-th/0606141} {arXiv:hep-th/0606141} \BibitemShut
  {NoStop}%
%%CITATION = HEP-TH/0606141;%%
\bibitem [{\citenamefont {Hamilton}\ \emph
  {et~al.}(2006{\natexlab{b}})\citenamefont {Hamilton}, \citenamefont {Kabat},
  \citenamefont {Lifschytz},\ and\ \citenamefont {Lowe}}]{Hamilton:2005ju}%
  \BibitemOpen
  \bibfield  {author} {\bibinfo {author} {\bibfnamefont {A.}~\bibnamefont
  {Hamilton}}, \bibinfo {author} {\bibfnamefont {D.~N.}\ \bibnamefont {Kabat}},
  \bibinfo {author} {\bibfnamefont {G.}~\bibnamefont {Lifschytz}}, \ and\
  \bibinfo {author} {\bibfnamefont {D.~A.}\ \bibnamefont {Lowe}},\ }\href
  {\doibase 10.1103/PhysRevD.73.086003} {\bibfield  {journal} {\bibinfo
  {journal} {Phys.Rev.}\ }\textbf {\bibinfo {volume} {D73}},\ \bibinfo {pages}
  {086003} (\bibinfo {year} {2006}{\natexlab{b}})},\ \Eprint
  {http://arxiv.org/abs/hep-th/0506118} {arXiv:hep-th/0506118} \BibitemShut
  {NoStop}%
%%CITATION = HEP-TH/0506118;%%
\bibitem [{\citenamefont {Hamilton}\ \emph {et~al.}(2007)\citenamefont
  {Hamilton}, \citenamefont {Kabat}, \citenamefont {Lifschytz},\ and\
  \citenamefont {Lowe}}]{Hamilton:2007wj}%
  \BibitemOpen
  \bibfield  {author} {\bibinfo {author} {\bibfnamefont {A.}~\bibnamefont
  {Hamilton}}, \bibinfo {author} {\bibfnamefont {D.~N.}\ \bibnamefont {Kabat}},
  \bibinfo {author} {\bibfnamefont {G.}~\bibnamefont {Lifschytz}}, \ and\
  \bibinfo {author} {\bibfnamefont {D.~A.}\ \bibnamefont {Lowe}},\ }\href@noop
  {} {\  (\bibinfo {year} {2007})},\ \Eprint {http://arxiv.org/abs/0710.4334}
  {arXiv:0710.4334} \BibitemShut {NoStop}%
%%CITATION = ARXIV:0710.4334;%%
\bibitem [{\citenamefont {Van~Raamsdonk}(2009)}]{VanRaamsdonk:2009ar}%
  \BibitemOpen
  \bibfield  {author} {\bibinfo {author} {\bibfnamefont {M.}~\bibnamefont
  {Van~Raamsdonk}},\ }\href@noop {} {\  (\bibinfo {year} {2009})},\ \Eprint
  {http://arxiv.org/abs/0907.2939} {arXiv:0907.2939} \BibitemShut {NoStop}%
%%CITATION = ARXIV:0907.2939;%%
\bibitem [{\citenamefont {Van~Raamsdonk}(2010)}]{VanRaamsdonk:2010pw}%
  \BibitemOpen
  \bibfield  {author} {\bibinfo {author} {\bibfnamefont {M.}~\bibnamefont
  {Van~Raamsdonk}},\ }\href {\doibase 10.1007/s10714-010-1034-0,
  10.1142/S0218271810018529} {\bibfield  {journal} {\bibinfo  {journal}
  {Gen.Rel.Grav.}\ }\textbf {\bibinfo {volume} {42}},\ \bibinfo {pages} {2323}
  (\bibinfo {year} {2010})},\ \Eprint {http://arxiv.org/abs/1005.3035}
  {arXiv:1005.3035} \BibitemShut {NoStop}%
%%CITATION = ARXIV:1005.3035;%%
\bibitem [{\citenamefont {Van~Raamsdonk}(2011)}]{VanRaamsdonk:2011zz}%
  \BibitemOpen
  \bibfield  {author} {\bibinfo {author} {\bibfnamefont {M.}~\bibnamefont
  {Van~Raamsdonk}},\ }\href {\doibase 10.1088/0264-9381/28/6/065002} {\bibfield
   {journal} {\bibinfo  {journal} {Class.Quant.Grav.}\ }\textbf {\bibinfo
  {volume} {28}},\ \bibinfo {pages} {065002} (\bibinfo {year}
  {2011})}\BibitemShut {NoStop}%
%%CITATION = CQGRD,28,065002;%%
\bibitem [{\citenamefont {Shenker}\ and\ \citenamefont
  {Stanford}(2014{\natexlab{a}})}]{Shenker:2013pqa}%
  \BibitemOpen
  \bibfield  {author} {\bibinfo {author} {\bibfnamefont {S.~H.}\ \bibnamefont
  {Shenker}}\ and\ \bibinfo {author} {\bibfnamefont {D.}~\bibnamefont
  {Stanford}},\ }\href {\doibase 10.1007/JHEP03(2014)067} {\bibfield  {journal}
  {\bibinfo  {journal} {JHEP}\ }\textbf {\bibinfo {volume} {1403}},\ \bibinfo
  {pages} {067} (\bibinfo {year} {2014}{\natexlab{a}})},\ \Eprint
  {http://arxiv.org/abs/1306.0622} {arXiv:1306.0622} \BibitemShut {NoStop}%
%%CITATION = ARXIV:1306.0622;%%
\bibitem [{\citenamefont {Shenker}\ and\ \citenamefont
  {Stanford}(2014{\natexlab{b}})}]{Shenker:2013yza}%
  \BibitemOpen
  \bibfield  {author} {\bibinfo {author} {\bibfnamefont {S.~H.}\ \bibnamefont
  {Shenker}}\ and\ \bibinfo {author} {\bibfnamefont {D.}~\bibnamefont
  {Stanford}},\ }\href {\doibase 10.1007/JHEP12(2014)046} {\bibfield  {journal}
  {\bibinfo  {journal} {JHEP}\ }\textbf {\bibinfo {volume} {1412}},\ \bibinfo
  {pages} {046} (\bibinfo {year} {2014}{\natexlab{b}})},\ \Eprint
  {http://arxiv.org/abs/1312.3296} {arXiv:1312.3296} \BibitemShut {NoStop}%
%%CITATION = ARXIV:1312.3296;%%
\bibitem [{\citenamefont {Mathur}(2014)}]{Mathur:2014dia}%
  \BibitemOpen
  \bibfield  {author} {\bibinfo {author} {\bibfnamefont {S.~D.}\ \bibnamefont
  {Mathur}},\ }\href@noop {} {\  (\bibinfo {year} {2014})},\ \Eprint
  {http://arxiv.org/abs/1402.6378} {arXiv:1402.6378} \BibitemShut {NoStop}%
%%CITATION = ARXIV:1402.6378;%%
\bibitem [{\citenamefont {Avery}\ and\ \citenamefont
  {Chowdhury}(2013)}]{Avery:2013bea}%
  \BibitemOpen
  \bibfield  {author} {\bibinfo {author} {\bibfnamefont {S.~G.}\ \bibnamefont
  {Avery}}\ and\ \bibinfo {author} {\bibfnamefont {B.~D.}\ \bibnamefont
  {Chowdhury}},\ }\href@noop {} {\  (\bibinfo {year} {2013})},\ \Eprint
  {http://arxiv.org/abs/1312.3346} {arXiv:1312.3346} \BibitemShut {NoStop}%
%%CITATION = ARXIV:1312.3346;%%
\bibitem [{\citenamefont {Kraus}\ \emph {et~al.}(2003)\citenamefont {Kraus},
  \citenamefont {Ooguri},\ and\ \citenamefont {Shenker}}]{Kraus:2002iv}%
  \BibitemOpen
  \bibfield  {author} {\bibinfo {author} {\bibfnamefont {P.}~\bibnamefont
  {Kraus}}, \bibinfo {author} {\bibfnamefont {H.}~\bibnamefont {Ooguri}}, \
  and\ \bibinfo {author} {\bibfnamefont {S.}~\bibnamefont {Shenker}},\ }\href
  {\doibase 10.1103/PhysRevD.67.124022} {\bibfield  {journal} {\bibinfo
  {journal} {Phys.Rev.}\ }\textbf {\bibinfo {volume} {D67}},\ \bibinfo {pages}
  {124022} (\bibinfo {year} {2003})},\ \Eprint
  {http://arxiv.org/abs/hep-th/0212277} {arXiv:hep-th/0212277} \BibitemShut
  {NoStop}%
%%CITATION = HEP-TH/0212277;%%
\bibitem [{\citenamefont {Fidkowski}\ \emph {et~al.}(2004)\citenamefont
  {Fidkowski}, \citenamefont {Hubeny}, \citenamefont {Kleban},\ and\
  \citenamefont {Shenker}}]{Fidkowski:2003nf}%
  \BibitemOpen
  \bibfield  {author} {\bibinfo {author} {\bibfnamefont {L.}~\bibnamefont
  {Fidkowski}}, \bibinfo {author} {\bibfnamefont {V.}~\bibnamefont {Hubeny}},
  \bibinfo {author} {\bibfnamefont {M.}~\bibnamefont {Kleban}}, \ and\ \bibinfo
  {author} {\bibfnamefont {S.}~\bibnamefont {Shenker}},\ }\href {\doibase
  10.1088/1126-6708/2004/02/014} {\bibfield  {journal} {\bibinfo  {journal}
  {JHEP}\ }\textbf {\bibinfo {volume} {0402}},\ \bibinfo {pages} {014}
  (\bibinfo {year} {2004})},\ \Eprint {http://arxiv.org/abs/hep-th/0306170}
  {arXiv:hep-th/0306170} \BibitemShut {NoStop}%
%%CITATION = HEP-TH/0306170;%%
\bibitem [{\citenamefont {Hartman}\ and\ \citenamefont
  {Maldacena}(2013)}]{Hartman:2013qma}%
  \BibitemOpen
  \bibfield  {author} {\bibinfo {author} {\bibfnamefont {T.}~\bibnamefont
  {Hartman}}\ and\ \bibinfo {author} {\bibfnamefont {J.}~\bibnamefont
  {Maldacena}},\ }\href {\doibase 10.1007/JHEP05(2013)014} {\bibfield
  {journal} {\bibinfo  {journal} {JHEP}\ }\textbf {\bibinfo {volume} {1305}},\
  \bibinfo {pages} {014} (\bibinfo {year} {2013})},\ \Eprint
  {http://arxiv.org/abs/1303.1080} {arXiv:1303.1080} \BibitemShut {NoStop}%
%%CITATION = ARXIV:1303.1080;%%
\end{thebibliography}%
\end{document}